\documentclass[fleqn,twoside]{article}
\usepackage[headings]{espcrc2}

\readRCS
$Id: espcrc2.tex,v 1.2 2004/02/24 11:22:11 spepping Exp $
\usepackage{graphicx}

\newcommand{\AmS}{{\protect\the\textfont2
  A\kern-.1667em\lower.5ex\hbox{M}\kern-.125emS}}

\newcommand{\pbarp}{\mbox{$\overline{p}p$}}    
\newcommand{\gevcc}{\mbox{$GeV/c^2$}}          
\newcommand{\ipb}{\mbox{$pb^{-1}$}}   
\newcommand{\ifb}{\mbox{$fb^{-1}$}}   

\newcommand{\jpsi}{\mbox{$J/\psi$}}        
\newcommand{\mupm}{\mbox{$\mu^+\mu^-$}}    

\title{Current Topics in D\O\ B-Physics}

\author{Arthur Maciel
        \address{Physics Department, Northern Illinois University,\\
          DeKalb, Illinois, 60115-2870 USA}
	\thanks{For the D\O\ Collaboration}}
       
\runtitle{Current Topics in D\O\ B-Physics}
\runauthor{Arthur Maciel}

\begin{document}

\begin{abstract}
An overview of selected topics in D\O\ B-Physics is presented, covering
relevant detector characteristics, and with emphasis on the most recent
results in the $B_s$, FCNC and rare decay programs.
\vspace{1pc}
\end{abstract}

\maketitle

\section{INTRODUCTION}

The Tevatron, Fermilab's \pbarp\ 1.96 \gevcc\ C.M. Collider, has recently
reached a delivered total of 1 \ifb\ of integrated luminosity to its two
multi-purpose detectors, D\O\ and CDF, during the current data taking
period known as RunII. Initial difficulties with beam stability were
efficiently corrected during a shutdown in late 2003, and since then
the Tevatron complex has operated remarkably well, with impressive
recovery during 2004 and matching its luminosity design goals in 2005
\cite{tevlum}. The D\O\ detector has been steadily accumulating good
quality data throughout RunII with luminosity recording efficiencies now 
approaching the 90\% mark \cite{d0lum}.

This article presents an overview of selected D\O\ studies
in B Physics, with emphasis on the most recent results, many of which are
still preliminary. Specific analyses are covered in more detail elsewhere 
in these proceedings.

\section{THE D\O\ DETECTOR}

D\O's central magnetic tracking volume is a key ingredient to the B Physics 
program. Inexistent during RunI, it is a RunII upgrade addition and consists
of a silicon microstrip tracker (SMT) surrounded by a central fiber tracker
(CFT), both inside a 2T superconducting solenoidal magnet. Optimized for
tracking and vertexing within $|\eta|<3$, the SMT has a six barrel
longitudinal structure, each with four layers of sensors. 
Barrels are interspersed with twelve F-disks (double sided, $\pm 15^o$
stereo), with four extra H-disks (single sided, $\pm 7.5^o$ stereo)
located beyond the barrels to complete the $\eta$ coverage. The CFT covers 
tracking within $|\eta|<1.6$, and has eight thin coaxial cylindrical layers, 
each supporting two doublets of overlapping scintillating fibers arranged 
either longitudinally along $z$ or in stereo pairs. Both the CFT and the SMT 
provide tracking information to the trigger.

Outside the coil the next two layers of detection consist of a preshower 
(scintillator strips) and a liquid argon calorimeter. There follows an
outer muon system, another key ingredient to D\O's B Physics program
because of its ability to efficiently trigger on semi-muonic $b$ decays
(or dimuons from $\jpsi$ etc),
thus providing a remarkable selection tool for $b$ events in the hostile
\pbarp\ collider environment. The muon system covers $|\eta|<2$ and
consists of a layer of multiple wire tracking devices and scintillation
counters in front of 1.8T iron toroids, followed by another two similar
detection layers outside the toroids. A detailed description of the
D\O\ detector can be found in \cite{d0nim}.

Every analysis presented here has one or more muons in its final state.
This illustrates the importance of muons in the D\O\ collection of a
B enriched sample. Trigger decisions are organized in a tiered system with
three levels of event inspection. Levels 1 and 2 B triggers are typically 
seeded by muon hits and complemented by tracking information. Level 1
muon candidates are about 60\% pure, and can be further purified at 
Level 2 with adjustable time gates and/or 3D track matches. The CFT
provides a fast and continuous readout to the Level 1 while the SMT
provides signals to the levels 2 and 3 \cite{sasha} for the detection
of displaced vertices from $b$-quark decays. Level 3 receives slower
more complete signals from all subdetectors enabling total reconstruction
of the event, with the ability for accurate tracking, vertexing and
invariant masses. Examples of B hadron reconstruction yields from the
D\O\ $\jpsi$ dimuon triggers are given in table \ref{tab:yields}.

\begin{table}[htb]
\caption{Number of decays reconstructed per \ipb.}
\label{tab:yields}
\newcommand{\m}{\hphantom{$-$}}
\newcommand{\cc}[1]{\multicolumn{1}{c}{#1}}
\renewcommand{\tabcolsep}{2pc} 
\renewcommand{\arraystretch}{1.2} 
\begin{tabular}{@{}lc}
\hline
decay    & $\sim N$/\ipb  \\ \hline
$B_u^\pm \rightarrow \jpsi + K^\pm$            & 21 \\
$B_d^0 \rightarrow \jpsi + K^\ast$  ($K\pi$)   &  8 \\
$B_d^0 \rightarrow \jpsi + K_s^0$   ($\pi\pi$) &  2 \\
$B_s^0 \rightarrow \jpsi + \phi$    ($KK$)     &  2 \\
$\Lambda_b^0 \rightarrow \jpsi + \Lambda^0$  ($p\pi$) & 0.3\\
\hline
\end{tabular}\\[2pt]
\end{table}

\section{THE $B_s$ PROGRAM}

One of the primary goals of RunII flavor physics at the Tevatron is
the observation of $B_s$ mixing and the extraction of $\Delta m_s$.
As part of its general effort in this direction, D\O\ presents in
this conference two complementary analyses of the $B_s$ system; (i)
a preliminary measurement of $\Delta\Gamma_s/\Gamma_s$ \cite{sanchez}
and (ii) an exercise towards the measurement of $\Delta m_s$ \cite{moulik}.

For access to the relative width difference between $B_s$ mass eigenstates
($\Delta\Gamma_s/\Gamma_s$) the decays $B_s \rightarrow \jpsi \phi$
are selected due to their non-flavor-specific final state, common to both
meson and anti-meson decays. Assuming
the two $B_s$ mass eigenstates correspond to the CP eigenstates, their
components in the end products are separated by means of a simultaneous
study of the $B_s$ time evolution, and the angular distribution of the final
state particles \cite{dgamma}. $\jpsi\phi$ states in a P (S or D) wave
indicate decays of the CP-odd (even) eigenstates. A {\em transversity}
angular variable particularly sensitive to this discrimination is defined.
An unbinned maximum likelihood fit to the data (about 450 \ipb) is performed,
that combines the candidate $B_s$ mass, lifetime and transversity. Three
parameters are extracted;
(i) $f_{odd}^{(t=0)}$, the relative rate of the CP-odd states at time
zero, (ii) $\bar\tau = 1/\bar\Gamma$ where 
$\bar\Gamma \equiv (\Gamma_{even} + \Gamma_{odd})/2$, and (iii)
$\Delta\Gamma/\bar\Gamma$.
The resulting values are shown in table \ref{tab:dgamma}.

\begin{table}[htb]
\caption{Results from the combined fit.}
\label{tab:dgamma}
\newcommand{\m}{\hphantom{$-$}}
\newcommand{\cc}[1]{\multicolumn{1}{c}{#1}}
\renewcommand{\tabcolsep}{2pc} 
\renewcommand{\arraystretch}{1.2} 
           {
             $
             \begin{array}{ccc}
               \hline
	       \mathrm{parameter} & \qquad & \mathrm{value} \\
               \hline
               \bar\tau & \qquad &
1.39 {\tiny \begin{array}{l}+0.13\\-0.14 \end{array} } \pm 0.08 ~ps \\
               \Delta\Gamma/\bar{\Gamma} & \qquad &
               0.21 {\tiny \begin{array}{l}+0.27\\-0.40 \end{array} }
               \pm 0.20 \\
               f_{odd}^{(t=0)}    & \qquad &   0.17 \pm 0.10 \pm 0.02 \\
               \hline
             \end{array}
             $
           }
\end{table}

Figure \ref{fig:dgamma} shows the one sigma contours (stat and stat+syst) 
for $c\bar\tau$ versus $\Delta\Gamma/\bar\Gamma$, overlaid with the 
result from CDF \cite{sanchez,cdf:dgamma} and the Standard Model prediction
\cite{sm:dgamma}.

\begin{figure}[htb]
\includegraphics[scale=0.38]{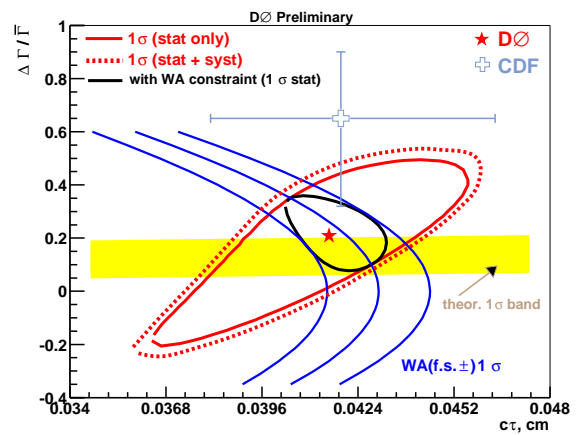}
\vspace{-1cm}
\caption{1$\sigma$ contours for $c\bar\tau$ versus 
$\Delta\Gamma/\bar\Gamma$.}
\label{fig:dgamma}
\end{figure}

A suite of single muon triggers offers access to a large sample of
semileptonic $B_s$ decays. With these, D\O\ has recently completed
what is currently the single most precise measurement of the $B_s$
lifetime \cite{taubs}. About 5000 $B_s \rightarrow D_s \mu X$ decay 
candidates were selected from 400 \ipb\ of data, where the $D_s$ is
reconstructed from its $\phi\pi$ final state. Ignoring lifetime 
differences ({\em i.e.} assuming the simple $e^{-t/\tau}$ as functional 
form) the value 
$\tau(B_s) = 1.420 \pm 0.043 \pm 0.057 ~\mathrm{ps}$
is obtained, and is compared with previous measurements and the revised
world average \cite{hfag} in figure \ref{fig:hfag}.

\begin{figure}[htb]
\includegraphics[scale=0.4]{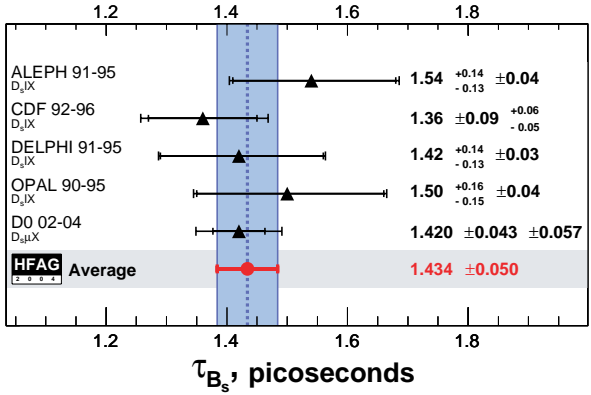}
\vspace{-1cm}
\caption{Individual $\tau(B_s)$ measurements and the world average
\cite{hfag}.}
\label{fig:hfag}
\end{figure}

This measurement of $\tau(B_s)$ provides an independent constraint
between the $c\bar\tau$ and $\Delta\Gamma/\bar\Gamma$ of figure
\ref{fig:dgamma} (since both results are extracted from different
data samples) and therefore can be used to further reduce the 
1$\sigma$ contours in that figure. This effect is represented in
figure \ref{fig:dgamma} by the three parabolas which map the 
uncertainty band of the WA value of figure \ref{fig:hfag} into
the $c\bar\tau$ versus $\Delta\Gamma/\bar\Gamma$ plane. The smaller
(inner) contour then represents the D\O\ results now under these tighter
constraints.

The large $B_s \rightarrow D_s \mu X$ decay sample, from which 
$\tau(B_s)$ is extracted, was also used in an exercise at setting
a lower limit to $\Delta m_s$, aimed at developing the tools and
methods for this measurement which still remains critically
dependent on the accumulation of luminosity. The initial state flavor
tagger builds a $b/\bar b$ likelihood from a choice of opposite side
discriminating variables, and 380 fully tagged events are selected in
the visible proper decay length range of $(-100,600)\mu$m, from which
the limit $\Delta m_s>5 ps^{-1}$ is extracted at 95\% CL. This analysis
is described in \cite{moulik,delms1}.

Since this conference, an updated version of the $\Delta m_s$ analysis
has been released \cite{delms2}. Some upgrade features are improved
selection methods and flavor taggers, plus an integrated luminosity
of 610 \ipb. $D_s \rightarrow KK^*$ is an added contribution, and
soft electron tagging is used for the first time in D\O, in combination
with other taggers. The currrent result is $\Delta m_s>7.3 ps^{-1}$
at 95\% CL, and indicates the pace at which this project progresses.
Large areas of improvement still remain untapped, covering analysis
techniques (added channels, same side tagging, unbinned fits)
as well as the use of hadronic $B_s$ decays.


\section{FCNC, RARE DECAYS}

D\O\ has search programs for {\em fcnc} decays in both $B$ and $C$ sectors.
Forbidden at tree level in the Standard Model (SM), these decays must
proceed via higher order {\em WI} processes. Such suppression allows for
physics beyond the SM, such as large $\tan\beta$ {\em SUSY} scenarios 
with an extended Higgs sector, to become competitive with, or even dominant
over the SM itself.

In the $B$ sector,  D\O's upper limit to $BR(B_s \rightarrow \mu^+\mu^-)$
\cite{bmumu} has recently been updated \cite{bmumunew} for additional
integrated luminosity, and set at $BR < 3.7\cdot 10^{-7}$ (95\% CL), still 
two orders of magnitude above SM expectations. A parallel analysis towards
setting an upper limit to $BR(B_s \rightarrow \mu^+\mu^-\phi)$ was work
in progress at the time of this conference. Its result has recently been
released \cite{bmumuphi} at $BR < 4.1\cdot 10^{-6}$ (95\% CL), now
approaching the SM predicted \cite{smbmuphi} central value of 
$1.6\cdot 10^{-6}$.

The $C$ sector is a rather different (and comparatively untapped)
testing ground for new phenomena. While the $B$ sector has 
{\em top}-assisted higher order loops, $C$ sector {\em fcnc} 
transitions involve
loops with lighter fermions and have a more efficient GIM cancellation.
SM extensions may upset this cancellation, and a sensitive observable 
\cite{gustavo} to which D\O\ has efficient access (via dimuon triggers)
is the differential decay rate ${\cal R} = d\Gamma(\mupm\pi)/dm(\mupm)$. 
D\O\ has started a program of searches for excess counts in ${\cal R}$,
in the continuum regions of the $\mupm\pi$ spectrum. To this end, a first 
step has been a study of resonant regions for setting normalization 
benchmarks. With 508 \ipb\ of analyzed data, D\O\ observes the
$D_s^+ \rightarrow \phi\pi^+ \rightarrow \mupm\pi^+$ final state with
a significance above background greater than 7$\sigma$ \cite{fcnc}.
The signal is shown in figure \ref{fig:fcnc} where a neighboring
peak suggests a 2.7$\sigma$ signal for the 
$D^+ \rightarrow \phi\pi^+ \rightarrow \mupm\pi^+$ decay, from which
a preliminary (and best existing) limit for this decay is set at
$BR < 3.1\cdot 10^{-6}$ (90\% CL).

\begin{figure}[htb]
\includegraphics[scale=0.4]{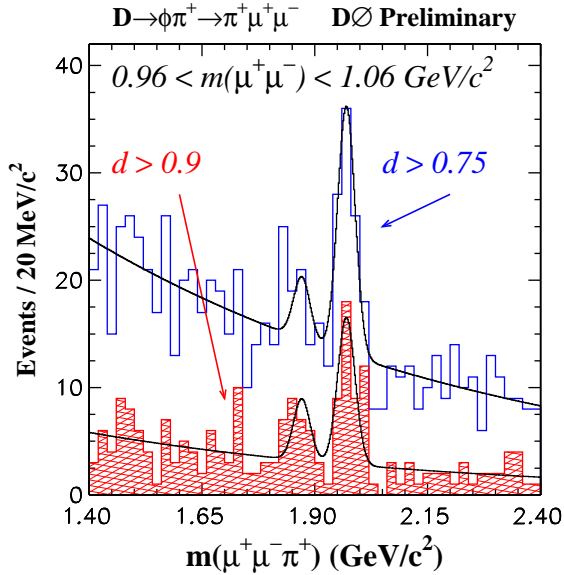}
\vspace{-1cm}
\caption{ The $\mupm\pi$ mass spectrum for two values of a 
likelihood variable $d$.}
\label{fig:fcnc}
\end{figure}


\section{CONCLUSION}

D\O\ is producing a wealth of heavy flavor results in RunII, with
significant impact on world averages and limits. Key measurements such
as $B_s$ mixing are on track, with uncertainty dominated by statistics,
not systematics.
Beyond the studies presented here, there are results in spectroscopy,
production, baryons, lifetimes etc, some of which were covered in
this conference \cite{others}. In the near future, a Tevatron shutdown 
is scheduled for three months, when D\O\ will undergo trigger upgrades,
with an extra 50Hz of bandwidth to tape dedicated to flavor physics. 
An inner layer (Layer-0) will be added to the silicon detector. 

\end{document}